\documentclass[showpacs,twocolumn,amssymb,amsfonts,amsmath,prb,aps]{revtex4}
\usepackage{longtable,graphicx,epsfig}

\begin{document}
\title{Microscopic View on Short-Range Wetting\\ at the Free Surface of the Binary Metallic Liquid
Gallium-Bismuth:\\
An X-ray Reflectivity and Square Gradient Theory Study.}

\author{Patrick~Huber}
\email{p.huber@mx.uni-saarland.de} \altaffiliation[present
address:]{ Fakult\"at f\"ur Physik und Elektrotechnik,
Universit\"at des Saarlandes, 66041 Saarbr\"ucken, Germany}

\affiliation{Department of Physics, Harvard University, Cambridge
MA 02138 (U. S. A.)}

\author{Oleg~Shpyrko}
\affiliation{Department of Physics, Harvard University, Cambridge
MA 02138 (U. S. A.)}

\author{Peter~Pershan}
\affiliation{Department of Physics, Harvard University, Cambridge
MA 02138 (U. S. A.)}

\author{Ben~Ocko}
\affiliation{Department of Physics, Brookhaven National Lab, Upton
NY 11973 (U. S. A.)}

\author{Elaine~DiMasi}
\affiliation{Department of Physics, Brookhaven National Lab, Upton
NY 11973 (U. S. A.)}

\author{Moshe~Deutsch}
\affiliation{Department of Physics, Bar-Ilan University, Ramat-Gan
52900 (Israel)}

\date{\today}

\begin{abstract}
We present an x-ray reflectivity study of wetting at the free
surface of the binary liquid metal gallium-bismuth (Ga-Bi) in the
region where the bulk phase separates into Bi-rich and Ga-rich
liquid phases. The measurements reveal the evolution of the
microscopic structure of wetting films of the Bi-rich,
low-surface-tension phase along different paths in the bulk phase
diagram. A balance between the surface potential preferring the
Bi-rich phase and the gravitational potential which favors the
Ga-rich phase at the surface pins the interface of the two demixed
liquid metallic phases close to the free surface. This enables us
to resolve it on an $\AA$ngstr\"om level and to apply a
mean-field, square gradient model extended by thermally activated
capillary waves as dominant thermal fluctuations. The sole free
parameter of the gradient model, i.e. the so-called influence
parameter, $\kappa$, is determined from our measurements. Relying
on a calculation of the liquid/liquid interfacial tension that
makes it possible to distinguish between intrinsic and capillary
wave contributions to the interfacial structure we estimate that
fluctuations affect the observed short-range, \textit{complete}
wetting phenomena only marginally. A \textit{critical} wetting
transition that should be sensitive to thermal fluctuations seems
to be absent in this binary metallic alloy.
\end{abstract}

\pacs{61.25.Mv, 61.30.Hn, 68.10.--m, 61.10.--i }

\maketitle

\section{Introduction}
The concept of a wetting transition that was introduced
independently by Cahn \cite{Cahn1977} and Ebner and Saam
\cite{Ebner1977} in 1977, has stimulated a substantial amount of
theoretical and experimental work\cite{DeGennes1985, Sullivan1986,
Dietrich1988, Law2001, Bonn2001}. Due to its critical character it
is not only important for a huge variety of technological
processes ranging from alloying to the flow of liquids, but has
the character of an extraordinarily versatile and universal
physical concept, which can be used to probe fundamental
predictions of statistical physics. For example, as shall be
depicted in more detail later on, it can be employed in the case
of wetting dominated by short-range interactions (SRW) in order to
test the predictions for the break-down of mean-field behavior and
the necessary ''transition'' to a renormalization group regime,
where this surface phenomenon is significantly affected by thermal
fluctuations.

A wetting transition occurs for two fluid phases in or near
equilibrium in contact with a third inert phase, e.g., the
container wall or the liquid-vapor interface. On approaching the
coexistence critical point the fluid phase that is energetically
favored at the interface forms a wetting film that intrudes
between the inert phase and the other fluid phase. In general,
this surface phenomenon is a delicate function of both the
macroscopic thermodynamics of the bulk phases and the microscopic
interactions.

Whereas one of the seminal theoretical works\cite{Ebner1977} on
the wetting transition gave a microscopic view on this phenomenon,
experimental results at this same level of detail were only
recently obtained through application of X-ray and neutron
reflection and diffraction techniques\cite{Plech2000,
Tostmann2000, Schlossman2002}. Moreover, almost all of these
experimental studies dealt with systems, like methanol-cyclohexane
and other organic materials, that are dominated by long-range
van-der-Waals interactions\cite{Bonn2001}. The principal
exceptions to this are the studies of the binary metallic systems
gallium-lead (Ga-Pb) \cite{Chatain1996}, gallium-thallium (Ga-Tl)
\cite {Shim2001}, and gallium-bismuth (Ga-Bi) \cite{Nattland1995},
for which the dominant interactions are short-range.

\begin{figure}[tbp]
\epsfig{file=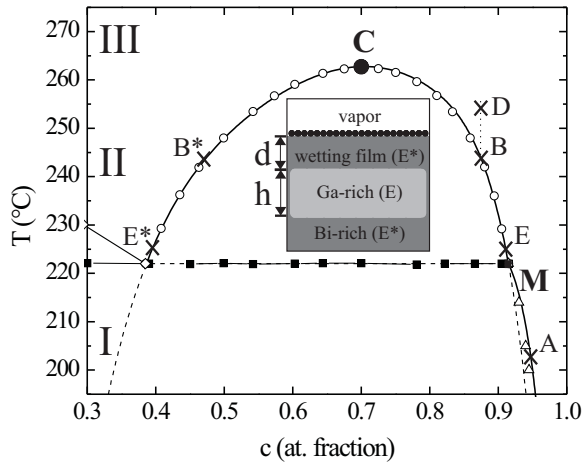, angle=0, width=1.00\columnwidth}
\caption{\label{fig:GaBiPDcT}The atomic fraction
($c$)-temperature($T$) bulk phase diagram of Ga-Bi. The symbols
indicate coexistence lines of Predel's phase
diagram\cite{Predel1960}. The lines show the phase boundaries
calculated from thermodynamic data\cite{Kaufman2001}. The dashed
lines represent the metastable extension of the (l/l) coexistence
line below $T_{\rm M}$. The points are: C-bulk critical point,
M-monotectic point, A,B,D,E-points on the experimental path. The
insets illustrate the surface and bulk phases. In region II the
wetting film is 50 \AA\ thick and the Ga-rich fluid is 5 mm thick.
The bold circless in the insets symbolize the Bi-monolayer. }
\end{figure}

Here, we will present an x-ray reflectivity study of wetting
phenomena that occur at the free surface in the binary metallic
liquid Ga-Bi when the bulk demixes in two liquid phases, i.e. a
Bi-rich and a Ga-rich liquid phase. The fact that the wetting
geometry pins the liquid/liquid (l/l) interface that divides the
two liquids to the surface allows measurements that show the
compositional profile of the wetting film at the $\AA$ngstr\"om
level. We will show results for the structure of the film as it
evolves towards the gravity-thinned film. Through a combination of
this structural information and the bulk thermodynamics of the
system, it was possible to extract detailed information on the
dominant interaction parameters governing the surface
phenomenology. We will show that a square gradient theory that is
combined with the effects of thermally excited capillary waves
provides a reasonable description of that interface. In fact, we
will be able to extract the sole free parameter of that model,
i.e. the influence parameter $\kappa$. This, in turn, will allow
us to distinguish between intrinsic (mean-field) and fluctuation
(capillary waves) contributions to the interfacial structure. On
the basis of this distinction we will estimate the influence of
fluctuations on the observed SRW at the free surface.


The paper is structured as follows: In the first section, we
introduce the bulk phase diagram of Ga-Bi and relate its topology
to the wetting transitions observable at the free surface of this
binary alloy. X-ray reflectivity measurements on the wetting films
along different paths in the bulk phase diagram will be discussed
in the second section. The third section will focus on the
thermodynamics and structure of the liquid-liquid interface. In
this section we will develop a square gradient theory in order to
model the concentration profile at the liquid/liquid interface.
Finally, in the last section we will provide a more general
discussion of the rich wetting phenomenology at the free surface
of this binary liquid metal.

\section{Bulk \& Surface Thermodynamics}
The bulk phase diagram of Ga-Bi, cp. Fig.~\ref{fig:GaBiPDcT},  was
measured by Predel with differential thermal
analysis\cite{Predel1960}. It is dominated by a miscibility gap
with consolute point C (critical temperature $T_C=262.8^{\circ}C$,
critical atomic fraction of Ga, $c_{\rm crit}=0.7$) and a
monotectic temperature, $T_M=222^{\circ}C$. In the part of region
I with $T<T_{\rm M}$, solid Bi coexists with a Ga-rich liquid. At
$T_{\rm M}$, the boundary between region I and region II, a first
order transition takes place in the bulk due to the liquidisation
of pure Bi. For $T_{\rm M}<T<T_{\rm C}$ (region II), the bulk
separates into two immiscible phases, a high density Bi-rich
liquid and a low density Ga-rich liquid. The heavier Bi-rich phase
is macroscopically separated from the lighter Ga-rich phase due to
gravity. In region III, beyond the miscibility gap, a homogeneous
liquid is found.


\begin{figure}[tbp]
\epsfig{file=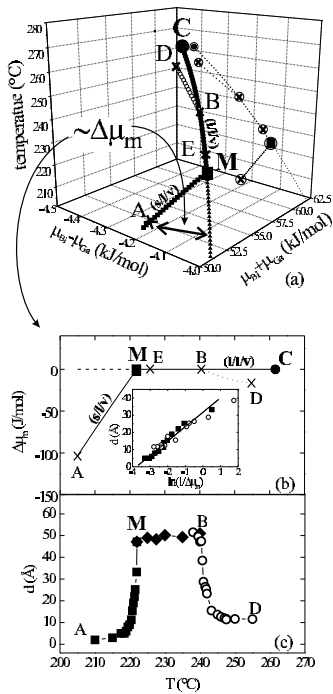, angle=0, width=0.5\columnwidth} 
\caption{\label{fig:GaBiPDmuT}\textbf{(a)} The chemical potential
($\mu$)-temperature ($T$) bulk phase diagram of Ga-Bi. The axes
are temperature, $T$, the difference $\left ( \mu_{\rm
Bi}-\mu_{\rm Ga} \right) $ of the chemical potentials of the
single species, and their sum $\left ( \mu_{\rm Bi}+\mu_{\rm Ga}
\right )$. The solid symbols indicate the following coexistence
lines: (solid circles) liquid/liquid/vapor (l/l/v) triple line,
(solid squares) solid/liquid/vapor (s/l/v) triple line, (small
triangles) metastable extension of the liquid/liquid/vapor (l/l/v)
triple line below $T_{\rm M}$, (open circles) experimental path
B-D probing complete wetting. The points are: C-bulk critical
point, M-monotectic point, A,B,D,E-points on the experimental
path. To illustrate the 3-dimensional structure of the phase
diagram, a projection of the phase boundary lines on the
($\mu_{\rm Bi}-\mu_{\rm Ga}$,$T$) plane is drawn in the plot. The
symbols surrounded by a circle show where the aforementioned
characteristic points fall on this projection.
\textbf{(b)}($\Delta \mu_{\rm m}$,$T$) phase diagram: (A-M) is the
(s/l/v), and (M-C) is the (l/l/v) coexistence line. The path B-D
is in the single phase region, and M and C are the monotectic and
critical points. Inset: effective wetting layer thickness $d$ on
A$\rightarrow $M (squares) and B$\rightarrow $D (open circles).
The solid line is a fit to the A$\rightarrow $M $d$-values.
\textbf{(c)} The measured $d$ along the experimental path.}
\end{figure}

Following the observation of Perepezko\cite{Perepezko1982} that on
cooling fine Ga-rich droplets are coated by a Bi-rich solid phase,
Nattland et. al. studied the liquid-vapor interface in region II
using ellipsometry. They found that a thin Bi-rich film intrudes
between the vapor and the Ga-rich subphase in defiance of gravity
\cite{Nattland1995}, as implied by the illustration in Fig.
\ref{fig:GaBiPDcT}. This was a clear example of the critical point
wetting in binary systems that was described by
Cahn\cite{Cahn1977}. On approaching to point C, the Bi-rich phase
necessarily becomes energetically favored at the free surface and
as a result it forms the wetting film that intrudes between the
Ga-rich subphase and the surface. In fact the situation is
slightly more complicated since x-ray studies indicate that
throughout region III the free surface is coated by a monolayer of
pure Bi\cite{Lei1996}. More recently we showed that on heating the
binary liquid a thicker wetting layer of Bi rich liquid forms
between the Bi monolayer and the bulk Ga rich liquid. This appears
to be an unusual example of complete wetting that is pinned to the
monotectic temperature $T_M$. This phenomenon was first discussed
by Dietrich and Schick in order to explain an analogous finding in
the binary metallic alloy Ga-Pb \cite{Chatain1996, Dietrich1997}.
The nature of this apparent coincidence of a surface transition
with the first order transition in the bulk at $T_M$ can most
easily be illustrated by a transformation of the ($c,T$)-diagram
to the appropriate chemical potential-temperature ($\mu$,
$T$)-diagram that is depicted in Fig. \ref{fig:GaBiPDmuT}(a) for
Ga-Bi. The axis are temperature, $T$, the difference $\left (
\mu_{\rm Bi}-\mu_{\rm Ga} \right) $ of the chemical potentials of
the single species, and their sum $\left ( \mu_{\rm Bi}+\mu_{\rm
Ga} \right )$. In this plot, the (l/l)-miscibility gap of Fig.
\ref{fig:GaBiPDcT}, which strictly spoken is a liquid/liquid/vapor
(l/l/v) coexistence boundary, transforms into a (l/l/v)-triple
line extending from $M$ to $C$ (solid circles). At M this triple
line intersects with another triple line, i.e. the
solid/liquid/vapor (s/l/v)-coexistence line (solid squares),
rendering M a \textit{tetra point} of four phase
coexistence\footnote{Additionally, the (s/l/v) triple line due to
the coexistence of a Bi-rich liquid, a pure Bi solid, and the
vapor phase - cp. the left side of the ($c$,$T$) phase diagram
(Fig. \ref{fig:GaBiPDcT}) starts at M. Since it is not relevant
for the considerations of the wetting thermodynamics at the
surface, it is not shown in Fig. \ref{fig:GaBiPDmuT}(a),(b) for
the sake of simplicity.}. At M a solid Bi, a Ga-rich , a Ga-poor
phase, and the vapor coexist.

The thermodynamics by which the surface transition at M is pinned
by the bulk phase transition is obvious if one considers the
topology of the ($\mu$,T)-plot in the proximity of M. The wetting
of the free surface by the Bi-rich phase as well as the bulk
transition are driven by the excess free energy, $\Delta\mu_{m}$,
of the Bi-rich phase over that of the Ga-rich liquid phase
\cite{Dietrich1988}. This quantity is proportional to the distance
between the (l/l/v)-triple line and any other line leading off
(l/l/v)-coexistence, e.g. the (s/l/v)-triple line
(A$\rightarrow$M) or the line B$\rightarrow$D in Fig.
\ref{fig:GaBiPDmuT}(a). The wetting thermodynamics is displayed in
a slightly simpler way by the plot $\Delta \mu_{\rm m}$ vs. $T$ in
Fig. \ref{fig:GaBiPDmuT}(b). In this figure for $T>T_{M}$, the
(l/l/v) coexistence line transforms into a horizontal straight
line that extends from M to C. For $T<T_{\rm M}$ the horizontal
dashed line indicates the metastable (l/l/v) extension of the
coexistence and that is above the solid-Bi/Ga-rich/vapor (s/l/v)
coexistence line that goes from M to A. This illustrates the
observation by Dietrich and Schick\cite{Dietrich1997} that the
path A$\rightarrow$ M leads to coexistence, and thus complete
wetting is dictated by the topology of the phase diagram.

A more quantitative understanding of the surface wetting phenomena
can be developed by analyzing the grand canonical potential,
$\Omega _{\rm S}$ per unit area A of the
surface\cite{Dietrich1988}: $\Omega _{\rm S}/A=d\,
\Delta\mu+\gamma_0\, e^{-d/\xi}$. Here, $d$ is the wetting film
thickness, $\xi$ is the decay length of a short-range, exponential
decaying potential, $\gamma_0$ its amplitude and $A$ an arbitrary
surface area. The quantity $\Delta\mu$ comprises all energies that
are responsible for a shift off true bulk (l/l/v) coexistence,
i.e. the aforementioned quantity $\Delta\mu_m$. The formation of
the heavier Bi-rich wetting layer at some height, $h$, above its
bulk reservoir costs an extra gravitational energy
$\Delta\mu_g$=$g\Delta\rho_m h$ where $\Delta\rho_{\rm m}$ is the
mass density difference between the two phases. Minimization of
$\Omega_{\rm S}$ in respect to $d$ then yields the equilibrium
wetting film thickness of the Bi-rich phase $d=\xi
\ln(\gamma_0/\Delta\mu)$. In fact the gravitational energy is only
significant in comparison with the other terms for very small
values $\Delta\mu_m$ and for most of the data shown in Fig.
\ref{fig:GaBiPDmuT} the gravitational term can be neglected,
allowing  of $\Delta\mu=\Delta\mu_m$. Thus, one expects a
logarithmic increase of the wetting film thickness upon
approaching M from A that is given by $d=\xi
\ln(\gamma_0/\Delta\mu_m)$, in agreement with the experimental
finding that was presented in our recent Physical Review Letter on
this phenomenon - cp. inset in Fig. \ref{fig:GaBiPDmuT}(b)
\cite{Huber2002}. The slight deviations for small values of
$\Delta\mu_m$, as well as the finite value of $d\sim50\AA$ along
the coexistence line M$\rightarrow$B are due to the gravitational
term. Since this approach of the (l/l/v)-coexistence line ends in
M with its four phase coexistence, the phenomenon is properly
described as \textit{tetra point wetting}. As demonstrated here,
the occurrence of this complete wetting phenomenon at the surface
is an intrinsic feature of the bulk phase diagram.

In this paper we shall present additional x-ray reflectivity
measurements that show the evolution of the wetting film on
approaching coexistence from point D in regime III, and hence
along a path probing complete wetting that is not dictated by the
topology intrinsic to the bulk phase diagram, but rather by the
experimentator's choice of the overall atomic fraction of Ga in
the sample $c_{\rm nom}$, i.e. the path B$\rightarrow$D in Fig.
\ref{fig:GaBiPDcT} and Fig. \ref{fig:GaBiPDmuT}(a,b). Furthermore,
we will probe the wetting film structure along an on-(l/l/v)
coexistence path B$\rightarrow$ E.

\section{X-ray reflectivity measurements}
\subsection{Experiment}

\begin{figure}[tbp]
\epsfig{file=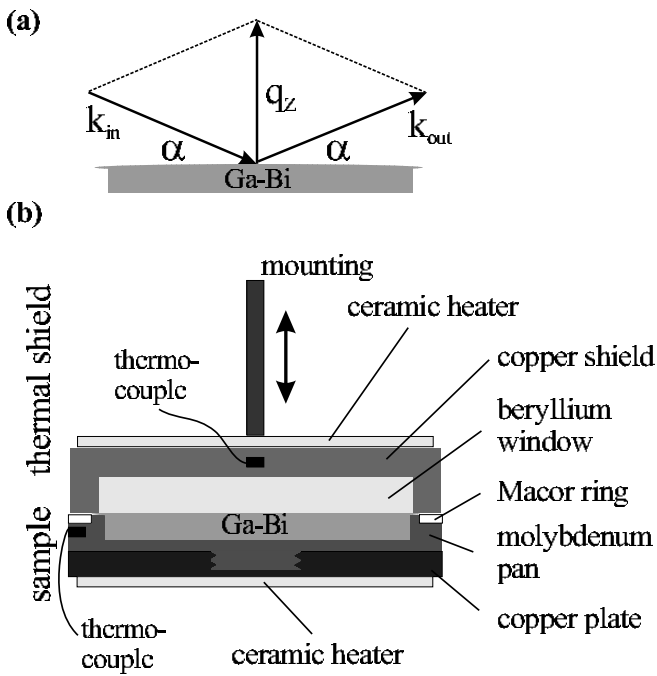, angle=0, width=1.00\columnwidth}
\caption{\label{fig:ExpSetup} (a) scattering geometry (b) Sketch
of the experimental setup. The arrow close to the mounting
indicates the variable position of the thermal shield in respect
to the sample surface.}
\end{figure}

\subsubsection{Sample Preparation \& Sample Environment}
The Ga-Bi alloy was prepared in an inert-gas box using
$>99.9999\%$ pure metals. A solid Bi pellet was covered by an
amount of liquid Ga required for a nominal concentration $c_{\rm
nom}=88$ at\% Ga. It was then transferred in air into an ultrahigh
vacuum chamber. A 24-hour period of bake-out yielded a pressure of
10$^{-10}$ torr. The residual surface oxide on the liquid's
surface was removed by sputtering with Ar$^{+}$ ions. Using
thermocouple sensors and an active temperature control on both
sample pan and its adjacent thermal shield a temperature stability
and uniformity of $\pm 0.05^{\circ }C$ was achieved. A sketch of
the experimental setup, sample environment resp. can be found in
Fig. \ref{fig:ExpSetup}(b).

A challenge in all x-ray reflectivity measurements of liquid
metals is the high surface tension of these systems, e.g. pure Ga
has a surface tension of $\approx$700mN/m. It often prevents
liquid metal from wetting non-reactive container walls or
substrates. This also leads to large curvatures of the liquid
surface hampering x-ray reflectivity measurements\cite{Regan1997}.
Remnant oxide layers that exists at the liquid/container,
liquid/substrate interface resp. even enhance this non-wetting
effect; therefore, we removed these oxide layers from the Mo
sample pan by sputtering with Ar$^{+}$ ions, at a sputter current
of 25mA and a sputter voltage of 2kV, and could reach wetting of
the Mo crucible by the liquid metal. This resulted in small
curvatures of the free surface as was judged by eye and later on
by x-ray reflectivity measurements. The resulting flat surfaces
facilitated the accumulation of reliable x-ray reflectivity data
sets, particularly for small incident angles, $\alpha$. On the
other hand, the wetting of the container wall by the liquid alloy
promoted some spilling of the liquid during the sample movements
that were necessary to track the sample position during the
reflectivity scan. We circumvented this problem by installing a
ceramic ring that surrounded the sample pan. The ceramics is not
wetted by the liquid metal and, therefore, inhibits any liquid
flow at the outermost circumference of the sample pan.

Another experimental challenge is related to x-ray reflectivity
measurements from liquids in general. The surface of a liquid is
sensitive to any kind of vibrations or acoustic noise. In order to
obtain a stable reflected signal from the liquid surface, we
mounted our UHV chamber on an active vibration
isolation\cite{Regan1997}.

\subsubsection{X-Ray Reflectivity}
X-ray reflectivity measurements were carried out using the liquid
surface reflectometer at beamline X22B at National Synchrotron
Light Source with an x-ray wavelength $\lambda =1.54\,$\AA .
Background and bulk scattering were subtracted from the specular
signal by displacing the detector out of the reflection plane by
$0.3 ^{\circ}$. The scattering geometry can be found as inset in
Fig. \ref{fig:ExpSetup}. The intensity $R(q_{\rm z})$, reflected
from the surface, is measured as a function of the normal
component $q_{\rm z} =4\pi / \lambda \sin(\alpha)$ of the momentum
transfer and yields information on the surface-normal structure of
the electron density $\rho (z)$ as described by the so-called
master formula\cite{Pershan1984}. This formula relates the
Fourier-transformed electron density gradient with the
experimentally obtained $R/R_{\rm F}$. The symbol $R_{\rm F}$
denotes the Fresnel reflectivity that is expected from an ideally
flat and sharp surface having the electron density of the Ga-rich
liquid. The standard procedure for determining the electron
density profile $\rho(z)$ from the measured reflectivity $R(q_{\rm
z})$ is to construct a simple and physically meaningful density
model and to fit its Fourier transform to experimentally obtained
data sets of $R/R_{\rm F}$\cite{Pershan1994}. We employ a
three-box model \cite{Tostmann2000}, where the upper box
represents the Gibbs-adsorbed Bi monolayer the second represents
the Bi-rich wetting film and the lower box represents the bulk
liquid. The relative electron densities of these boxes correspond
to pure Bi (top box) and to the electron density of the Bi-rich
wetting film, $\rho$. The quantity $\rho_{\rm sub}$ denotes the
electron density of the Ga-rich subphase. In the simplest
approximation the electron density profiles of the interfaces
between the different phases can be described analytically by
error-functions. The relevant model for this problem includes
three error-functions (erf)-diffusenesses for the following
interfaces: vapor/Bi monolayer, Bi monolayer/Bi rich film, Bi rich
film/Ga rich bulk phase. The first two interfaces describing the
monolayer feature were remained unchanged in the presence of the
Bi-rich film, while only the diffuseness of the Bi rich film/Ga
rich bulk phase interface, $\sigma_{\rm obs}$, was a variable
parameter as a function of $T$. The model also included size
parameters that describe the thickness of the two upper boxes. The
box that describes the bulk sub-phase extends to infinity. During
the fitting the thickness of the monolayer box was kept constant
while the thickness of the Bi-rich film was allowed to vary.

\begin{figure}[tbp]
\epsfig{file=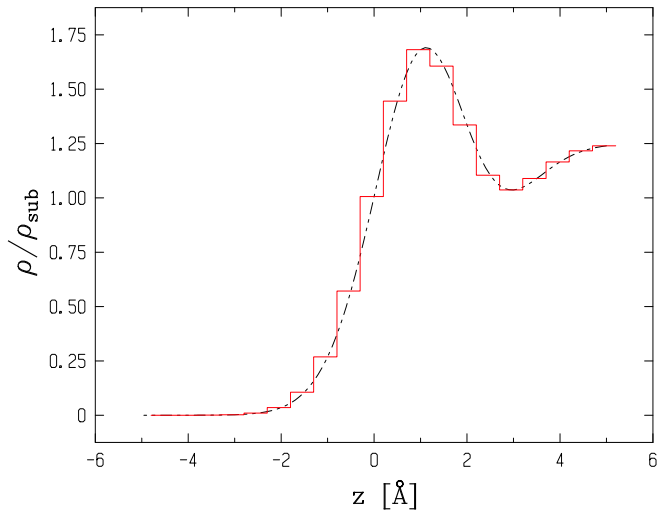, angle=0, width=1.00\columnwidth}
\caption{\label{fig:Parratt} Illustration of the approximation of
the analytic electron density profile (dashed line) by thin slabs
in order to apply the Parratt formalism. Here, only the region
close to the monolayer feature is depicted.}
\end{figure}

The use of the master formula tacitly assumes the validity of the
Born approximation. This assumption holds true for wave vectors,
$q_{\rm z}$, much larger than the value of the critical wavevector
$q_{\rm c}$ of the Ga-rich subphase, where multi-scattering
effects can be neglected\cite{Tolan1998}. Here, however, we will
be interested in features corresponding to 30-60 $\AA$ thick
wetting films, which when translated to momentum-space corresponds
to features in $q_{\rm z}$ around 0.05-0.1 ${\rm \AA}^{-1}$ that
are comparable to the value of the critical wavevector $q_{\rm c}
\approx 0.05  {\rm \AA}^{-1}$ (The q$_c$-values change only
slightly depending on $T$, cp. Table \ref{tab:FitsAB}). In order
to deal with this we resort to the recursive Parratt formalism
\cite{Parratt1954} for $q_{\rm z}$-values close to $q_{\rm c}$. In
this formalism one develops a 2x2 matrix that relates the
amplitudes and phases for the incoming and outgoing waves on both
sides of a slab of arbitrary dielectric constant. For the present
problem we approximate the above mentioned three-box profile by a
large number of thinner slabs whose amplitudes are chosen to
follow the envelope of the analytic profile - cp. fig.
\ref{fig:Parratt}. Typically the number of slabs was of the order
of 300.

\subsection{Wetting film structures approaching (l/l/v)-coexistence}
\begin{figure}[tbp]
\epsfig{file=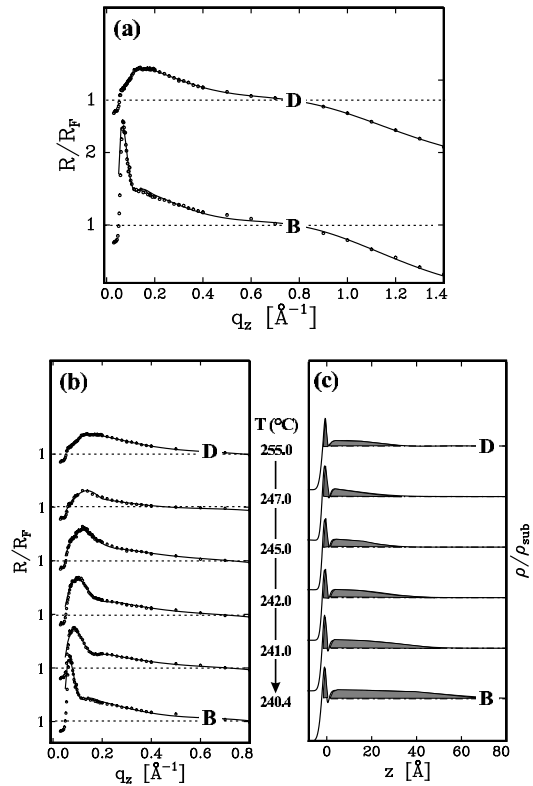, angle=0, width=0.8\columnwidth} 
\caption{\label{fig:RRFDensPathBD}\textbf{(a)} Normalized
reflectivities $R/R_{\rm F}$ for selected points D and B
corresponding to $T_{\rm D}$=255.0$^{\circ}$C, $T_{\rm
B}=240.4^{\circ}$C. Dashed lines indicate $R/R_{\rm F}=1$.
\textbf{(b)} $T$-dependent normalized reflectivities $R/R_{\rm F}$
while approaching coexistence on path D $\rightarrow$B along with
fits. Dashed lines indicate $R/R_{\rm F}=1$.\textbf{(c)} Electron
density profiles $\rho$/$\rho_{\rm sub}$. All regions with
$\rho/\rho_{\rm sub}>1$ indicating Bi-enrichment as compared with
the Ga-rich subphase are gray shaded.}
\end{figure}

Here we will present x-ray reflectivity measurements from the free
surface of Ga-Bi along a path approaching (l/l/v) coexistence from
the homogeneous part of the bulk phase diagram (regime III), i.e.
the path D$\rightarrow$B . This path, particularly the position of
its characteristic point B, is solely determined by the overall
atomic fraction of Ga in the sample pan, $c_{\rm nom}$. While
cooling this path intersects the (l/l/v) triple line (cp. Fig.
\ref{fig:GaBiPDmuT}) at the point B, corresponding to temperature
$T_{\rm B}$ of the miscibility boundary (cp. Fig.
\ref{fig:GaBiPDcT}). For $T<T_{\rm B}$ the heretofore homogeneous
bulk liquid phase separates into the heavier Bi rich liquid that
settles towards the bottom of the pan and the lighter Ga rich
liquid on top. The actual position in the ($c$,$T$)-plane was
chosen from consideration of the vanishing electron density
contrast between Ga-rich and Bi-rich phase upon approaching C on
the one hand or the (l/l/v) triple line (cp. Fig.
\ref{fig:GaBiPDmuT}(a)) enforcing the spinodal demixing of the
afore homogeneous bulk liquid. In our case, its position in the
($c$,$T$)-plane was chosen after a consideration of the vanishing
electron density contrast between Ga-rich and Bi-rich phase upon
approaching C, on the one hand, and a reasonably ''long path'' on
(l/l/v) coexistence on the other hand. The compromise was a value
of about $12\%$ for c$_{\rm nom}$, which corresponds to  $T_{\rm
B}=240.4^{\circ}C$.

X-ray reflectivity $R(q_{\rm z})$ was measured at selected
temperatures on path D$\rightarrow$B. In order to avoid artifacts
associated with the fact that the evolution of the wetting film is
governed by slow diffusion processes \cite{Lipowsky1986, Wu1986,
Fenistein2002} equilibration was monitored by continuous taking
repeated reflectivity scans following increments of small
temperature steps of $0.5^{\circ}C$. Typically the measured
reflectivity fluctuated wildly for a couple of hours, after which
it slowly evolved to a stable equilibrium. The fits (lines) to
these ''equilibrium'' $R/R_{\rm F}$ (points) are shown in
Fig.\ref{fig:RRFDensPathBD}(a,b), and the corresponding $\rho (z)$
profiles - in Fig.~\ref{fig:RRFDensPathBD}(c).  At point D
[T$_{\rm D}=255^{\circ}C$], typical of region III, $R/R_{\rm F}$
exhibits a broad peak at low $q_{\rm z}$ as well as an increased
intensity around $q_{\rm z}=0.8\AA^{-1}$. The corresponding
electron density profile $\rho(z)$ obtained from the fit indicates
a thin, inhomogeneous film of increased electron density (as
compared with $\rho_{\rm sub}$) close to the surface along with a
peak right at the surface. It is consistent with a segregated
monolayer of pure Bi along with a thin layer of a Bi-enriched
phase. As the temperature is decreased towards B, the peak is
gradually shifting to lower $q_{\rm z}$ and its width decreases.
This behavior manifests the continuous growth in thickness of the
wetting layer upon approaching $T_{\rm B}$ and is in agreement
with the thermodynamic path probing complete wetting:
$\Delta\mu_{\rm m}\rightarrow0$ on path D$\rightarrow$B.

Our experimental data clearly indicate film structures dominated
by sizeable gradients in the electron density, which contrasts
with frequently used "homogeneous slab" models, but is in
agreement with theoretical calculations. These range from density
functional calculations via square gradient approximations to
Monte Carlo simulations for wetting transitions at hard walls
\cite{Ebner1977, Davis1996, Schmid2001}. Inhomogeneous profiles
have also been observed experimentally in microscopically resolved
wetting transitions for systems dominated by long-range
van-der-Waals interactions \cite{Plech2000, Plech2001}. Clearly
detailed interpretation of the inhomogeneity for the GaBi wetting
films will require either a density functional analysis, or some
other equivalent approach. Nevertheless, even a simple model
approximating the wetting layer by a slab of thickness $d$ allows
a reliable determination of the surface potential governing this
complete wetting transition. In order to do so, effective film
thicknesses $d$ have been extracted form the $\rho(z)$ profiles.
In Fig. \ref{fig:GaBiPDmuT}(c) we show plots of these $d$-values
versus both $T$ and $\Delta\mu_{\rm m}$. The last plot shows the
expected logarithmic behavior and allows determination of the
values for the parameters of the short-range potential, i.e.
$\gamma_{\rm 0}=400mN/m$ and $\xi=5.4\AA$ \cite{Huber2002}.
Moreover, agreement between the new ($d$,$\mu_m$)-behavior along
$D \rightarrow B$ and the data for the path A$\rightarrow$M
provides an experimental prove that both paths have the same
thermodynamic character, i.e. they probe complete wetting:
$\Delta\mu_{\rm m}\rightarrow 0 $ on path A$\rightarrow$M and
D$\rightarrow$B.

\subsection{Wetting film structures: On (l/l/v)-coexistence}
\begin{figure}[tbp]
\epsfig{file=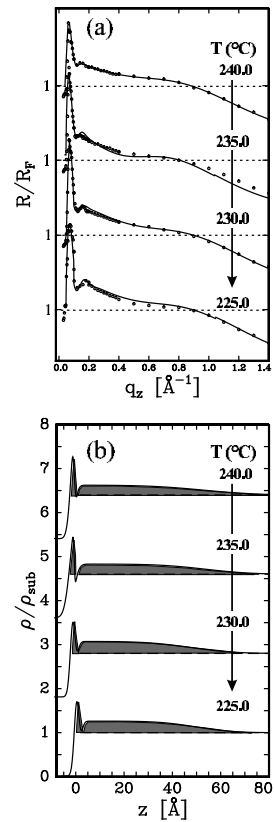, angle=0, width=0.4\columnwidth} 
\caption{\label{fig:RRFDensAB}\textbf{(a)} T-dependent normalized
reflectivities R/R$_{\rm F}$ while leaving coexistence on path B
$\rightarrow$D along with fits. Dashed lines indicate $R/R_{\rm
F}=1$. With increasing $T$, each $R/R_F$ is shifted by $1.2$.
\textbf{(b)} Electron density profiles $\rho/\rho_{\rm sub}$. All
regions with $\rho/\rho_{\rm sub}>1$ indicating Bi-enrichment as
compared with the Ga-rich subphase are gray shaded.}
\end{figure}

As discussed before, on cooling from the one phase region (III) we
reach the (l/l/v)-coexistence at B. The data for the measured
dependence of $R/R_{\rm F}$ on $q_{\rm z}$ along coexistence
between points B and M are shown in Fig. \ref{fig:RRFDensAB}(a).
The existence of two peaks at low $q_{\rm z}$ indicates the
presence of a fully formed thick film ($\sim 50\AA$). The solid
lines indicate the best fit results corresponding to the real
space profiles that are shown in Fig. \ref{fig:RRFDensAB}(b). The
best fit value for maximum density of the thick film $\rho /\rho
_{\rm sub}=1.20$ agrees reasonably well with the value of 1.21
that is calculated form the phase diagram at point B.

These results are also reasonably consistent with what is expected
for a gravity limited slab of uniform density. Using $\xi=5.4\AA$,
$\sigma_0=400mN/M$ and the known material constants that make up
$\mu_m$ (see table \ref{tab:FitsAB} below) the calculated value
for $d=d_{\rm g}=\xi \ln (\sigma_{\rm 0}/\Delta\mu_{\rm
g})=15.6\xi=85\AA$. In view of the fact that this estimate does
not take into account the excess energy associated with associated
with concentration gradients across the interfaces some
overestimation of $d_{\rm g}$ is not too surprising. Nevertheless,
this rough calculation does show that the wetting film thickness
is expected to be on a mesoscopic rather than on the macroscopic
length scale that has been observed for similar wetting geometries
in systems governed by long-range, dispersion
forces\cite{Law2001}.

Upon cooling from point B to M, the intensity of the first peak of
$R/R_{\rm F}$ increases as expected due to the gradual increase of
the electron density contrast $\rho/\rho_{\rm sub}$ that follows
from the increase in the Bi concentration of the bulk phase.
Similarly, there is a small shift in the position of the peak
indicating that the thickness varies slightly from $53\AA$ near B
to $50\AA$ near E that follows from the $T$-dependence of the
density contrast $\Delta\rho_{\rm m}(T)$ as estimated from the
bulk phase $\Delta\rho_{\rm m}(T)$ - cp. Table \ref{tab:FitsAB}.
The on-coexistence path B$\rightarrow$E is too far away from C to
expect more pronounced effects on the wetting layer thickness due
to a vanishing density contrast or the increase of the influence
of the criticality on the interaction potentials
\cite{Dietrich1988, Bonn2001, Bonn2002}.

Moreover the gravitational thinning of the on-coexistence wetting
film thickness to a length scale comparable to the range of the
exponentially decaying short-range interactions prevents testing
the influence of long-range, van-der-Waals like atomic
interactions on the wetting behavior that is present regardless of
whether the system is insulating or metallic\cite{Ashcroft1991,
Pluis1989}.

Finally, we would like to highlight the unique wetting geometry
encountered here: The subtle balance between the surface potential
that favors the Bi-rich liquid phase at the surface and the
gravitational potential that favors the Ga-rich liquid phase above
the denser Bi-rich phase, pins the (l/l) interface between the two
coexisting phases close to the free surface. It is this property
that makes it possible to study the structure of these wetting
films. For example, the width of the interface between that film
and bulk, $\sigma_{\rm obs}$, is the case of the decay of the
Kiessig fringes at low $q_{\rm z}$. This will be discussed further
in the following section.

\begin{table*}
\squeezetable
\begin{ruledtabular}
\begin{tabular}{ccccccccccc}
  $T (^{\circ} C)$ &$c^{\rm I}$ &$c^{\rm II}$ &$\Delta \rho$ $_{\rm m}$ &$\rho/\rho_{\rm sub}(calc)$ &$\rho/\rho_{\rm sub}(obs) $&q$_c$ &d$_g$ &$d$ &$\sigma_{\rm obs}$ &$\sigma_{\rm calc}$ \\
  \hline
  225.0 &0.39& 0.91& 2.30 &1.24 &1.25 &0.0498 & 84.7 &50 &11.78&12.0 \\
  \hline
  230.0 &0.41& 0.90& 2.20 &1.24 &1.25 &0.0499 & 84.9 &51 &14.2 &13.3 \\
  \hline
  235.0 &0.43& 0.89& 2.09 &1.22 &1.22 &0.0500 & 85.2 &52 &15.0 &14.9 \\
  \hline
  240.0 &0.45& 0.88& 1.95 &1.21 &1.21 &0.0502 & 85.5 &53 &17.5 &17.25 \\
\end{tabular}
\end{ruledtabular}
\caption{\label{tab:FitsAB}Material parameters of the coexisting
liquid phases as calculated from the bulk phase diagram as well as
electron density profiles as obtained from our fits to $R/R_{\rm
F}$ for selected temperatures $T$ along the on-coexistence path
B$\rightarrow$E. The atomic fraction of Ga in the coexisting
Ga-rich, Bi-rich liquid phase are $c^{\rm I}$, $c^{\rm I}$ resp.
The symbols $\rho/\rho_{\rm sub}(calc)$, $\rho/\rho_{\rm
sub}(obs)$ denote the calculated, observed electron density ratios
resp.}
\end{table*}

\section{Liquid/Liquid Interface} In this section, we will discuss the microscopic structure
of the (l/l)-interface separating the gravitationally thinned film
of the Bi-rich phase from the Ga-rich subphase along the
on-coexistence path B$\rightarrow$E. We shall first describe a
simple square gradient theory for this interface and then extend
it by the consideration of thermally excited capillary waves. We
will use this theory in order to extract the (l/l)-interfacial
profile and the (l/l)-interfacial tension from our measurements of
the average surface structure.

\subsection{Square Gradient Theory}
At the (l/l)-interface between the two phases, the concentration
varies continuously from the bulk concentration of the homogeneous
Bi-rich phase, $c^{\rm I}$, to the bulk concentration of the
homogeneous Ga-rich phase, $c^{\rm II}$. Assuming that the
variation in concentration within the interface is gradual in
comparison with the intermolecular distance, the excess free
energy for the inhomogeneous region can be expanded in terms of
local variables $c(\overrightarrow{r})$ and $\bigtriangledown
c(\overrightarrow{r})$\cite{Davis1996}. The Gibbs free energy
density cost within the interface can then be expressed in terms
of a combination of a local function $g(c(\overrightarrow{r},T)$
and a power series in $\bigtriangledown c(\overrightarrow{r})$
\cite{VanderWaals1894, Cahn1958, Safran1994,Davis1996}:
\begin{eqnarray}
\widetilde{G}=N\int_{\rm V}\left[ g(c(\overrightarrow{r}),T)+
\frac{1}{2}\kappa \left(\nabla c(\overrightarrow{r})
\right)^{2}+... \right] dV \label{eq:GTaylor}
\end{eqnarray}
where $N$ is the number of molecules per unit volume. The function
$g(c(\overrightarrow{r},T)$ is the Gibbs free energy density that
a volume would have in a homogeneous solution. The next term is
the leading term in the power series expansion. The coefficient
$\kappa$ referred to as the influence parameter characterizes the
effect of concentration gradients on the free energy. From first
principles, it is related to the second moment of the
Ornstein-Zernike direct correlation function and can be replaced
by the pair potential weighted mean square range of intermolecular
interactions of the system\cite{Davis1996}. Equation
(\ref{eq:GTaylor}) can be considered the Landau-Ginzburg
functional for this problem.

Applying (\ref{eq:GTaylor}) to the one-dimensional composition
change $c(z)$ across the interface and neglecting terms in
derivatives higher than the second gives for the total Gibbs free
Energy $\widetilde{G}$ of the system:
\begin{eqnarray}
\widetilde{G}=NA_{\rm 0}\int_{-\infty}^{+\infty} \left[ g(c(z)) +
\frac{1}{2}\kappa \left( \frac{dc(z)}{dz} \right)^2 \right]dz
\label{eq:GradientApprox}
\end{eqnarray}
The variable $z$ represents the direction perpendicular to the
interface. The symbol $A_{\rm 0}$ denotes an arbitrary surface
area. The interfacial tension $\gamma_{\rm ll}$ is defined as the
excess energy of this inhomogeneous configuration in respect to
the Gibbs free energy $G$ of the homogeneous liquid of one of the
coexisting phases. The reference system we have chosen is the
homogeneous Bi-rich liquid with concentration $c^{\rm I}$:
\begin{eqnarray}
\gamma_{\rm ll}=\frac{1}{A_{\rm 0}}\left[
\widetilde{G}(c(z)-G(c^{\rm I}) \right] \label{eq:SigmaDef}
\end{eqnarray}
Combining the gradient approximation (\ref{eq:GradientApprox})
with the thermodynamic definition of the interfacial tension
(\ref{eq:SigmaDef}) yields:
\begin{eqnarray}
\gamma_{\rm ll} = N \int_{-\infty} ^{+\infty} \left[ \Delta
g(c(z))+\frac{1}{2}\kappa \left(\frac{dc(z)}{dz}\right)^{2}\right]
dz
\end{eqnarray}
where the grand thermodynamic potential $\Delta g(c)$ is given by
\begin{eqnarray}
\Delta g(c(z))=g(c(z))-g(c^{\rm I})
\end{eqnarray}
It follows then from (\ref{eq:SigmaDef}) that the interfacial
tension is a functional of the concentration profile $c(z)$ at the
interface. The equilibrium surface tension is obtained by
minimizing this functional leading to the following Euler-Lagrange
equation:
\begin{eqnarray}
\Delta g(c)=\kappa\;\frac{d^2c(z)}{dz^2} \label{eq:DiffProfile}
\end{eqnarray}
This differential equation along with the boundary conditions that
the concentration has to vary from $c^{\rm I}$ on the one side of
the interface to $c^{\rm II}$ on the other side determines $c(z)$
unambiguously. In fact, the simplicity of (\ref{eq:DiffProfile})
allows its direct integration, yielding an expression for $z(c)$:
\begin{eqnarray}
z(c)=z_0+ \int_{c^{\rm I}}^{c^{\rm{II}}}
\sqrt{\frac{\kappa}{\Delta g(c)}}\; dc
\label{eq:DiffProfileQuadratur}
\end{eqnarray}
where $z_{\rm 0}$ and $c_{\rm 0}$ represent an arbitrary chosen
origin and composition. It should be noted, that from a
renormalization of the integrand of equation
(\ref{eq:DiffProfileQuadratur}), one can conclude that any
characteristic length scale, such as the intrinsic width of the
interfacial profile, $\sigma_{\rm intr}$, has to scale as
$\sqrt{\kappa}$.

An attractive feature of this gradient theory is that it relates
the profile to $\kappa$ and $g(c(\overrightarrow{r},T)$ without
regard for the theoretical basis by which
$g(c(\overrightarrow{r},T)$ is derived. In this paper we will use
the extended regular solution model that will be presented in the
appendix to model the homogeneous Free Energy of the binary liquid
metal, particularly the one used to model its miscibility gap. The
sole quantity required then in order to calculate the (l/l)
interfacial profile and tension is an expression for the influence
parameter $\kappa$, that precedes the square
gradient\cite{Enders1998}. It follows that the influence parameter
can be extracted from the measurement of the microscopic structure
of the interface. For example, from an appropriate choice for
$\kappa$, solving of (\ref{eq:DiffProfile}) or
(\ref{eq:DiffProfileQuadratur}) will yield a profile characterized
by the intrinsic width $\sigma_{\rm intr}$ that can be compared
with $\sigma_{\rm obs}$.

\subsection{Capillary Wave Excitations on the (l/l) Interface}
The formalism presented so far relies on a simple mean-field
picture that neglects any thermal fluctuations on the (l/l)
interface. By contrast, the real experiment is sensitive to not
only the mean-field diffuseness of the interface, but is also
sensitive to the fact that thermal fluctuations contribute
additional broadening to the fluid interface. An intuitive,
semiphenomenological approach to handle the interplay of these two
quantities was initiated by Buff, Lovett, and Stillinger
\cite{Buff1965, Rowlinson1982}. Essentially, they imagined the
interface as if it were a membrane in a state of tension
characterized by the bare interfacial energy $\gamma_{\rm ll}$, as
calculated in the former paragraph; it sustains a spectrum of
thermally activated capillary waves modes whose average energy is
determined by equipartition theorem to be $k_{\rm B}T/2$. By
integration over the spectrum of capillary waves, one finds the
average mean square displacement of the interface, $\sigma_{\rm
cap}$, as
\begin{eqnarray}
\sigma_{\rm cap}^2=\frac{k_{\rm B}T}{4 \pi \gamma_{\rm
ll}(\kappa)} \ln \frac{q_{\rm min}}{q_{\rm max}}
\label{eq:WidthCapillary}
\end{eqnarray}
Here $q_{\rm max}$ is the largest capillary  wave vector that can
be sustained by the interface. For bulk liquids this is typically
of the order of $\pi$/molecular diameter; however, it does not
seem realistic to describe excitations with wavelengths that are
smaller than the intrinsic interfacial width as surface capillary
waves{\cite{Wadewitz1999, Evans1981}. Thus $\sigma_{\rm intr}$
describes the intrinsic interfacial width $q_{\rm
max}=\pi/\sigma_{\rm intr}$. The value of the quantity $q_{min}$
is slightly more complicated since it depends on whether the
resolution is high enough to detect the long wavelength limit at
which an external potential $v(z)$, such as either gravity or the
van der Waals interaction with the substrate,  quenches the
capillary wave spectrum.  If the resolution is sufficiently high
then $q_{\rm min}=\pi/L_{\rm v}$ where $L_{\rm v}^2=\gamma_{\rm
ll}^{\rm -1}\frac{\partial^2 v(z)}{\partial z^2}$. If it is not
then the value of $\sigma_{\rm cap}^{\rm 2}$ is limited by the
resolution limited average length scale $L_{\rm res}$ over which
the observed fluctuations can be resolved. In this case $q_{\rm
min}=\Delta q_{\rm res}=\pi/L_{\rm res}$ where $\Delta q_{res}$ is
the projection of the detector resolution on the plane of the
interface. In our case $q_{\rm res}$ is significantly larger than
$q_{\rm v}$ therefore, $q_{\rm min}=q_{\rm res}=0.04\AA^{-1}$.

Using the fact that capillary waves obey a gaussian statistics
which manifests itself as an erf-function type profile for the
average interfacial roughness\cite{Daillant1999} together with the
observation that the intrinsic profiles $c(z)$ are also well
described by erf-function profiles, the two contributions to the
interfacial width can be added in gaussian
quadrature\cite{Daillant1999}. Thus we write the total observed
interfacial width, $\sigma_{\rm calc}$, as
\begin{eqnarray}
  \sigma_{\rm calc}^{2}=\sigma_{\rm intr}(\kappa)^2+ \sigma_{\rm cap}(\gamma_{\rm ll}(\kappa))^2
\label{eq:WidthQuadrature}
\end{eqnarray}

Comparison of (\ref{eq:WidthCapillary}) and (\ref{eq:DiffProfile})
reveals that both, $\sigma_{\rm intr}$ and $\sigma_{\rm cap}$
depend on $\kappa$. Thus, the determination of $\kappa$ leads to
the problem of finding the self-consistent value that yields a
value for $\sigma_{\rm calc}$ that agrees with the experimentally
determined roughness $\sigma_{\rm obs}$.

\subsection{Intrinsic Profile and (l/l) interfacial tension of
Ga-Bi}

We performed a self consistent calculation of $\kappa$ for the
observed (l/l) interface at point E, where $\sigma_{\rm
obs}(E)\approx12\AA$. In order to do so, we solved the
Euler-Lagrange Equation (\ref{eq:DiffProfile})\footnote{We used
the standard procedures for solving differential equations
included in the \protect \textit {Maple 7} software package
(Waterloo Maple Inc.) as well as the \protect \textit  {"Shooting
Technique for the solution of two-point boundary value problems"}
as implemented for this software package by D.B. Maede\cite
{Maede1996}.} which yields the intrinsic $c(z)$ profile as
depicted and compared with an erf-profile in Fig.
\ref{fig:LiqLiqIntrins}(a). From that we calculated the
corresponding erf-diffuseness $\sigma_{\rm intr}(E)$ and via
numerical integration of (\ref{eq:SigmaDef}) the corresponding
interfacial tension $\gamma(E)$ as well as the resultant capillary
induced roughness and the overall average diffuseness,
$\sigma_{\rm cap}$ and $\sigma_{\rm intr}$ resp. After starting
with a reasonable value of $\kappa$, as judged by a scaling
analysis of (\ref{eq:DiffProfile}), this procedure was iterated
until $\sigma_{\rm calc}(E)=\sigma_{\rm obs}(E)$. The value we
obtained was $\kappa_{\rm opt}=5.02\cdot10 ^{-13}Nm^3/mol$. The
other quantities of interest at E are: $\sigma_{\rm
cap}(E)=10.32\AA$ , $\sigma_{\rm intr}(E)=6.35\AA$ and
$\gamma_{\rm ll}(E)=3.3mN/m$.

It is interesting to observe that (\ref{eq:DiffProfile}) implies
$\sigma_{\rm intr}\propto \frac {1}{\kappa}$, whereas from
(\ref{eq:WidthCapillary}) one obtains $\sigma_{\rm cap} \propto
\sqrt{\kappa}$. Hence, the function of $\sigma_{\rm calc}(\kappa)$
exhibits a minimum for $\sigma_{\rm calc}$ at a characteristic
value, $\kappa_{\rm min}$. It is interesting to note, that the
$\kappa_{\rm opt}$, determined from our measurements, corresponds
to this peculiar value $\kappa_{\rm min}$ meaning that the system
selects from all possible $\sigma_{\rm intr}$, $\sigma_{\rm cap}$
combinations, the one with the minimum overall width for the
(l/l)-interface.

Assuming a $T$-independent influence parameter along with the
available thermochemical data sets that are available over a wide
range of the phase diagram allows calculation of the $T$-dependent
interfacial profiles and (l/l) interfacial tension from very low
up to very high $T$, close to $T_C$. In Fig.
\ref{fig:LiqLiqIntrins}(b) the intrinsic (l/l) interfacial
profiles are plotted for selected $T$'s versus the reduced
temperature, $t=|T-T_{\rm C}|/T_{\rm C}$ (Here $T$ denotes the
absolute temperature in Kelvin units). The increasing width of the
interface while approaching $T_C$ ($t\rightarrow0$) is
well-demonstrated. In Fig. \ref{fig:gammaGaBi} the $t$-dependent
interfacial tension, $\gamma_{ll}(T)$, is depicted. As expected,
it vanishes as $t\rightarrow0$.
\begin{figure}[tbp]
\epsfig{file=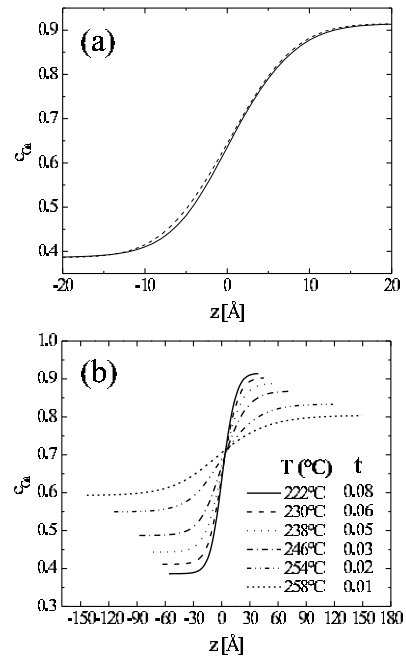, width=0.6\columnwidth} 
\caption{\label{fig:LiqLiqIntrins} \textbf{(a)} Calculated
intrinsic profile $c(z)$ ($-$) for $T=T_E=225^{\circ}C$, $t=0.07$
in comparison with the erf-function profile (dashed line):
$c(z)=\frac{c^{\rm I}+c^{\rm II}}{2}+\frac {(c^{\rm II}-c^{\rm
I})}{2}$ erf $\left ( \frac{z}{\sqrt{2} \sigma_{\rm intr}(E)}
\right)$ with $\sigma_{\rm intr}(E)=6.35\AA$. \textbf{(b)}
Calculated intrinsic concentration profiles $c(z)$ for selected
temperatures $T$, reduced temperatures $t$ resp. as indicated in
the figure.}
\end{figure}

\begin{figure}[tbp]
\epsfig{file=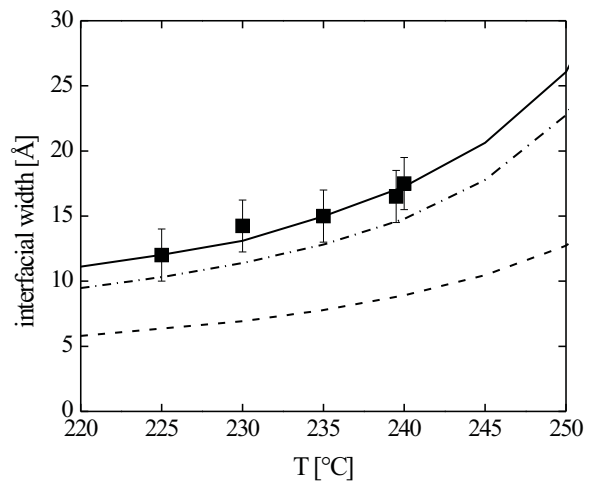, width=1.00\columnwidth}
\caption{\label{fig:CpCMW}Comparison of temperature dependent,
measured interfacial width $ \sigma $ ($\blacksquare$) with
calculated interfacial widths (lines): $\sigma _{\rm calc}$ =
$\sqrt{\sigma_{\rm intr}^{2}+\sigma_{\rm cap}^{2}}$ ($-$) ,
intrinsic width $\sigma_{\rm intr}$ ($\cdot-$), and capillary
width $\sigma_{\rm cap}$ ($--$).}
\end{figure}


Comparison of $\gamma_{\rm ll}(T)$ with other measurements would
be a useful test of our measurements and interpretation.
Unfortunately, measurements are rather difficult and we are not
aware of any such measurements for Ga-Bi. Fortunately, it is still
possible to estimate $\gamma_{\rm ll}(T)$ using other available
experimental data sets on this alloy. For example,
Predel\cite{Predel1960} noted that in the vicinity of C the
(l/l)-coexistence curve of Ga-Bi can be represented by a function
of the form $|\phi_{\rm c}-\phi|=K(T_{\rm c}-T)^{\rm \beta}$ with
$\beta\cong0.33$, where $\phi$ is the volume fraction of Ga, Bi
resp. and $K$ a constant \cite{Predel1960}. This behavior suggests
that the demixing transition in Ga-Bi belongs to the same
universality class as binary nonmetallic liquid mixtures, i.e.
Ising lattice model with dimensionality D=3. Moreover, it allows
for an estimation of the interfacial tension based on a scaling
relation for these quantities, i.e. the hypothesis of
two-scale-factor universality (TSFU) \cite{Stauffer1972} should
apply. Using this concept along with available $T$-dependent
measurements of the specific heat, Kreuser and Woermann extracted
an expression for the $T$-dependent interfacial tension in Ga-Bi
\cite{Kreuser1993}, $\gamma_{\rm ts} = \gamma_{\rm ts0}~t^{\rm
\mu}$, where $\mu$ is a critical exponent with universal character
for a bulk demixing transition, $\mu \cong 1.26$ and $\gamma_{\rm
ts0}=66\pm15mN/m$. The corresponding $t$-dependent $\gamma_{\rm
ts}$ ($-$) is plotted in Fig. \ref{fig:gammaGaBi}. A good
agreement with the $t$-dependence of $\gamma_{\rm ll}$ calculated
by our square gradient theory is found, e.g. at $t_{\rm E}=0.07,
T_{\rm E}=225^{\circ}C$: $\gamma(E)= 3.31mN/m$,
$\gamma_{ts}(E)=5.7mN/m$ \footnote{In the strict sense, the TSFU
hypothesis is only valid in a $T$-range, where mapping of the
demixing process on a second order phase transition is justified,
which is typically the case for reduced temperatures, $t<10^{\rm
-1}$. Despite of this restriction, it often fits well experimental
results\cite{Merkwitz1998} and obviously also our results of the
gradient theory in a much wider temperature range.}


\begin{figure}[tbp]
\epsfig{file=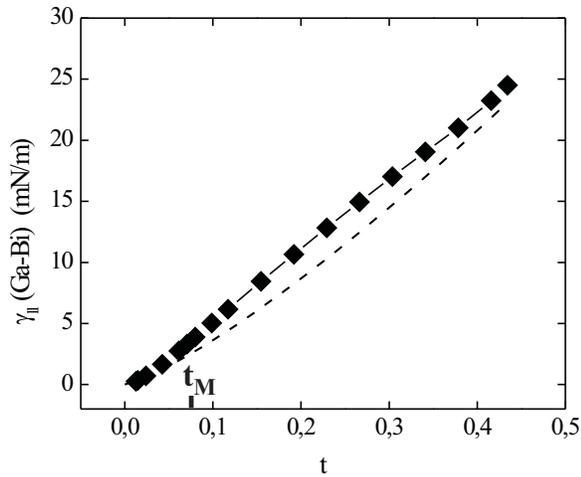, width=1.00\columnwidth}
\caption{\label{fig:gammaGaBi} Calculated (l/l) interfacial
tension $\gamma_{\rm ll}$ of Ga-Bi as a function of the reduced
temperature $t$. The dashed line represents the TSFU prediction
for Ga-Bi\cite{Kreuser1993}. The symbol $t_{\rm M}$ indicates the
monotectic temperature at $t_{\rm M}$(Ga-Bi)$=0.08$.}
\end{figure}

\subsection{(l/l) interfacial tension of Ga-Pb}

Aside from the Ga-Bi system the only other metallic system for
which detailed temperature dependent
measurements\cite{Merkwitz1998} and estimations
\cite{Wynblatt1998} of the (l/l) interface are reported, is Ga-Pb
. Given the close relationship between Ga-Bi and Ga-Pb (similar
constituents, identic phase diagram topology with consolute point
C and monotectic point M), we felt encouraged to also apply our
gradient theory to this binary system. Moreover, we assumed that
the influence parameter, $\kappa$, as determined by the
measurements on Ga-Bi, also provides a reasonable value for this
metallic system. Available thermochemical data sets for the
miscibility gap of Ga-Pb\cite{Ansara1991} allowed us to calculate
the $t$-dependent (l/l) interfacial tension. As can be seen in
fig. \ref{fig:gammaGaPb}, where our calculated values for
$\gamma_{\rm ll}$ are shown in comparison with the experimental
values, an excellent agreement is found.

\begin{figure}[tbp]
\epsfig{file=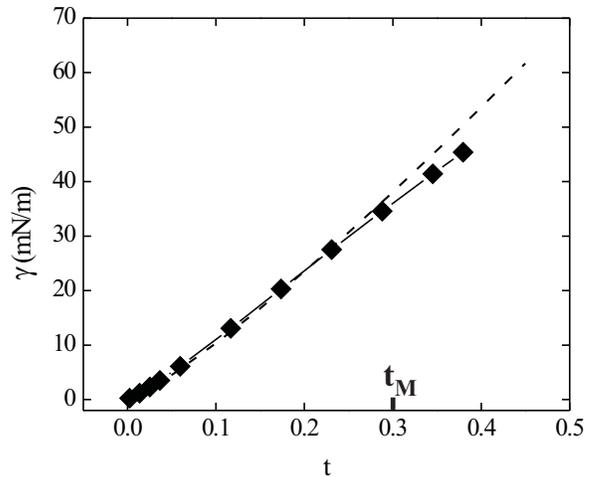, width=1.00\columnwidth}
\caption{\label{fig:gammaGaPb} Calculated (l/l) interfacial
tension $\gamma_{\rm ll}$ of Ga-Pb as a function of the reduced
temperature $t$. The solid points are our calculated values. The
dashed line represents a fit to the measurements by Merkwitz et.
al. \cite{Merkwitz1998}. The symbol $t_{\rm M}$ indicates the
monotectic temperature at $t_{\rm M}$(Ga-Pb)$=0.3$.}
\end{figure}

Overall, our results suggest that our model of an intrinsic,
mean-field interface broadened by capillary waves describes the
(l/l) interface reasonably well. One must, however, bear in mind
the approximation implicit in the presented analysis. For
instance, we have assumed that the free surface of the liquid
alloy behaves like a rigid wall, whereas it will have its own
spectrum of capillary waves that could be coupled to the waves at
the interface. However, since the interfacial tension of the free
surface is about two orders of magnitude larger than the (l/l)
interfacial energy in this metallic system, we believe that the
rigid wall assumption is a reasonable starting point. Furthermore,
the calculation presented above has the shortcoming of a
$T$-independent influence parameter $\kappa$.
It should be mentioned, though, that theory predicts a divergence
of $\kappa$ close to $T_{\rm C}$ for real fluids,
$\kappa\approx(T-T_{C})^{\rm -0.02}$ for $T\rightarrow T_{\rm C}$.
In fact, it can even be shown that this divergence is necessary to
obtain the correct scaling behavior of $\gamma_{\rm ll}(T)$ near C
\cite{Cornelisse1997, Fisk1969, Sengers1989}.

\section{Wetting Phenomenology at the Free Surface}
\subsubsection{Effects of Fluctuations on Short-Range Wetting}
Our analysis in section 1 of the complete wetting transition at B
was based on a mean-field (MF) model for a SRW transition. Such a
model accurately predicts the critical behavior (i.e. the critical
exponents) of systems close to a phase transition, provided
fluctuations can be neglected. For SRW transitions, however, it
can be shown that the upper critical dimensionality
$D_u=3$\cite{Schick1990}. The value $D_u$ is the dimension beyond
which MF theory can be applied successfully. If the dimension is
smaller than $D_u$ fluctuations are important and one has to
resort to renormalization group (RG) methods to describe the
critical behavior of a phase transition. Because its upper
critical dimensionality is exactly equal to 3, the SRW transition
has received a great deal of theoretical attention, since it
allows one to explore the regime where the MF behavior breaks down
due to fluctuations, and the RG approach becomes applicable. This
break-down depends on the so-called fluctuation parameter,
$\omega=k_{\rm B} \cdot T /(4\pi\gamma_{\rm ll}\xi_{\rm b}^2)$
(where $\xi_b$ is the bulk correlation length, and $\gamma_{\rm
ll}$ is the (l/l) interfacial tension) which measures the
magnitude of the dominant thermal fluctuations, the thermally
induced capillary waves at the fluid, unbinding interface, in our
case at the (l/l) interface of the coexisting Bi- and Ga-rich
liquids.

For a complete wetting transition RG analysis of the divergence of
the wetting film thickness yields a changed prefactor of the
logarithmic divergence, $d_{\rm RG}\sim\xi_{\rm RG}\,(1+\omega
/2)\, \ln\, (1 /\Delta \mu)$\cite{Dietrich1988, Brezin1983,
Kroll1985}. An estimation of $\omega$ using the bulk correlation
length, $\xi_b$, estimated from the aforementioned TSFU
prediction, yields for the wetting transitions at M, $\omega_{\rm
M}=0.3\pm0.2$ and for the observed wetting complete wetting
transition at B, $\omega_{\rm B}=0.4\pm0.2$. Thus, the prefactors
of the logarithmic increase of the wetting film with $\Delta\mu$
is changed by only $\approx$ 10\%, 20\% resp. in comparison with
the mean-field prediction. This is well within our experimental
error of about 30\% for the determination of $\xi$. (The large
error bar on $\xi$ is due to its extraction from a slab thickness
analysis whereas the measured wetting films have significant
inhomogeneous structures). A clear distinction between RG and MF
behavior cannot be drawn in our case.

A \textit{critical} wetting transition characterized by the
wetting film formation on-(l/l/v)-coexistence, while changing $T$,
with a similar value of $\omega$ should show more pronounced
deviations from MF behavior. Thus, we shall estimate the
characteristic temperature $T_{\rm W}$ of that transition in the
following.

\subsubsection{The Critical Wetting Transition in Ga-Bi}
It follows from our measurements that the critical wetting
transition has to be hidden in the metastable range of the (l/l/v)
coexistence line somewhere below $T_{\rm M}$. An estimation of
$T_{\rm W}$ is possible by a consideration of the Free energy at
the surface for the wet (Bi-rich wetting at surface) and non-wet
(Ga-rich phase at the surface) situation. In order to do so, we
introduce the spreading energy, $\Theta(T)$. It estimates the
preference of the Bi-rich phase in respect to the Ga-rich phase at
the free surface at any temperature and is given by the following
energetic contributions. In the wet-situation, the energy is given
by the sum of the free surface tension of the Bi-rich phase and
the tension of the (l/l) interface between the Bi-rich phase and
the Ga-rich subphase. The non-wet situation is characterized by
the surface energy of the bare Ga-rich liquid phase,
$\gamma(Ga-rich)_{\rm lv}$. We define $\Theta(T)$ therefore in the
following way:
\begin{eqnarray}
   \Theta(T) &=&\gamma(Ga-rich)_{\rm lv}(T)-\nonumber \\
             & &\left ( \gamma(Bi-rich)_{\rm lv}(T)+\gamma_{\rm ll}(T) \right )\nonumber \\
             &=&\left ( \gamma(Ga-rich)_{\rm lv}(T)-\gamma(Bi-rich)_{\rm lv}(T) \right )\nonumber \\
             & &-\gamma_{\rm ll}(T)\nonumber \\
          &=&\Gamma(T)-\gamma_{\rm ll}(T)
\label{eq:SpreadingEnergy}
\end{eqnarray}
In equation (\ref{eq:SpreadingEnergy}) the quantity $\Gamma(T)$
denotes the difference in the bare liquid-vapor surface tensions
of the coexisting phases.

From the definition of $\Theta(T)$ it follows that establishment
of the wetting film is energetically preferred if $\Theta>0$
(wet-situation), whereas for $\Theta<0$ it is unfavorable (non-wet
situation). The condition $\Theta(T)=0$ marks, therefore, the
transition temperature $T_{\rm W}$, where the system switches
on-coexistence from a non-wet to a wet-situation or vice versa.

In order to estimate $\Theta(T)$, in addition to $\gamma_{\rm
ll}(T)$, as extracted in the previous section we need measurements
or estimations of $\gamma(Ga-rich)_{\rm lv}(T)$ and
$\gamma(Bi-rich)_{\rm lv}(T)$. Measurements of these quantities
are reported in the literature\cite{Khokonov1974, Ayyad2002a,
Ayyad2002b}, the one by Ayyad and Freyland that employed the
noninvasive method of capillary wave spectroscopy is the most
recent one. From these measurements, we get a conservative
$T$-independent estimate for $\Gamma$ of the order of 100mN/m. Our
calculation of $\gamma_{\rm ll}(T)$, particularly its
extrapolation towards very low temperature $t\rightarrow1$ which
yields $\gamma_{\rm ll}(0K)=75mN/m$, indicates that it is always
significant smaller than $\Gamma$. Our analysis suggests,
therefore, that $\Theta(T)>0$ for all $T$ meaning that \textit{no}
critical wetting transition occurs on the metastable extension of
the (l/l/v) coexistence line.

This remarkable conclusion is, on the one hand, in contrast to the
findings for the few examples of binary metallic wetting systems
studied so far: In Ga-Pb and Ga-Tl, Wynblatt and Chatain et al.
estimated a value for $T_{\rm W}$ significanttly below the
corresponding monotectic temperatures T$_{\rm M}$, but still at
final values of the order of $0.3 \cdot T_C(K)$, $0.2 \cdot
T_C(K)$ for Ga-Pb, Ga-Tl resp. ($T_{\rm C}(K)$ denotes the
critical temperature of bulk demixing on the Kelvin temperature
scale.) On the other hand, it reflects the general finding that in
binary metallic systems with short-range interactions the critical
wetting transition occurs at significantly lower temperatures than
this is the case in organic liquid systems, where $T_{\rm W}$ is
found to lie above $0.5 \cdot T_{\rm C}(K)$. This tendency seems
to be driven to its extreme in the case of Ga-Bi: $T_W=0 \cdot
T_{\rm C}(K)=0K$.

A semiquantitative argument for the absence of a critical wetting
transtion in Ga-Bi as compared to the two other metallic systems
investigated so far might relate to the fact that in comparison
with the others the liquid/liquid coexistence region for GaBi
system is significantly shifted towards the Ga-rich, high surface
tension phase and, in addition the miscibility gab is much
narrower.  As a result the change in concentration across the l/l
interface is smaller for the GaBi system than for the others. In
view of the fact that one can express
\begin{eqnarray}
   \gamma_{\rm ll} = \kappa \int dz \frac{dc(z)}{dz}^{\rm 2}\approx \sqrt{\kappa}\left ( \Delta c \right )^2
\label{eq:GammaApprox}
\end{eqnarray}
this suggests that $\gamma_{ll}$ could be expected to be smaller
for GaBi than for the others. If this effect is important, and if
it is not compensated by an accompanying reduction in the value of
$\Gamma$, then it is possible that
$\Theta(T)=\Gamma-\gamma_{ll}>0$ for all temperatures, thereby
precluding a wetting transition.

\subsubsection{Tetra Point Wetting $\Leftrightarrow$ Surface Freezing}
Finally, we would like to comment on the fact that the observed
wetting by the Bi rich liquid at the GaBi monotectic point is not
necessarily the only phenomena that can occur. In principle one
expects at this tetra point a similar or even more diverse wetting
phenomenology as has been predicted \cite{Pandit1983} and found
\cite{Zimmerli1992} in the proximity of a triple point for one
component systems. For example, another possibility is that the
free surface could be wet by solid Bi. In the bulk both solid Bi
and the Bi rich liquid are stable for $T>T_M$; however,  for
$T<T_{\rm M}$ the free energy of the Bi rich liquid is larger than
that of solid Bi. Consequently, if the solid Bi surface phase had
a favorable spreading energy it would certainly wet the free
surface for $T<T_{\rm M}$. In fact, on the basis of optical and
surface energy measurements Turchanin et al. \cite{Turchanin2001,
Turchanin2002} proposed a surface phase diagram for which a
surface frozen" phase of solid Bi forms at the free surface of the
GaBi liquid for temperatures below $T_M$. Unfortunately, we have a
problem with this. The first may be purely semantic, but  the idea
of a "surface frozen phase" originates in the experiments by
Earnshaw et al. and Deutsch, Gang, Ocko et al \cite{Earnshaw1992,
Wu1993} on the appearance of solid surface phases at temperatures
above that of bulk freezing for alkanes and related compounds. For
the GaBi system the crystalline phase of bulk Bi is stable at
temperatures well above those of at which Turchanin et al made
their observations. We believe that if their observations do
correspond to  thermal equilibrium phenomena, the effect would be
more appropriately described as wetting of the free surface by
solid Bi. Unfortunately, the second problem we have is that we do
not believe that surface wetting by solid Bi is favored by the
spreading energy. If wetting by solid Bi were favored, as implied
by Turchanin et al, then the wetting layers of Bi rich liquid that
are the subject of this paper would have to be metastable,
existing only because of a kinetic barrier to the nucleation of
the solid wetting film. In fact we have regularly observed   that
on cooling below $T_{\rm M}$ the original fluid, smooth, highly
reflecting surfaces became both rough and rigid. The effect is
exacerbated by the presence of temperature gradients and by rapid
cooling. We interpreted this as the formation of bulk solid Bi,
rather than wetting,  since we never saw any direct evidence for
the formation of films, rather than bulk Bi.  We argued  that if
there not enough time for the excess Bi in the bulk liquid phase
below the surface to diffuse, and precipitate as bulk solid Bi at
the bottom of the liquid pan, the bulk liquid would simply undergo
bulk spinodal  phase separation. This is consistent with our
observations. Unfortunately it is very difficult to absolutely
prove that  our films are stable and this issue can't be  resolved
from the existing evidence.  Nevertheless we find it difficult to
understand why solid Bi would wet the surface for $T<T_{\rm M}$
and not wet the surface as $T$ approaches $T_{\rm M}$ from above.

\section{Summary}
In this paper we reported x-ray reflectivity measurements of the
temperature dependence of wetting phenomena that occur at the free
surface of the binary metallic alloy Ga-Bi when the bulk demixes
into two liquid phases, i.e. a Bi-rich and a Ga-rich liquid phase.
We characterized the temperature dependence of the thickness and
interfacial profile, with $\AA$ngstr\"om resolution,  of the
Bi-rich wetting films that form at the free surface of the liquid
on approaching the (l/l/v) coexistence triple line,  i.e. along a
path of complete wetting .  The results, which are characterized
by large concentration gradients, agree with density functional
calculations for such transitions at hard walls. On the basis of
this it was possible to determine the short-range surface
potential that favors the Bi-rich phase at the free surface and
hence is governing the wetting phenomenology. The short-range
\textit{complete wetting} transition turns out to be only
marginally affected by thermal fluctuations.  According to
renormalization group analysis the \textit{critical wetting}
transition should be more sensitive to critical fluctuations;
however, according to our analysis the wetting layer of Bi rich
liquid should be present for all temperatures. This is in contrast
to the results for the other binary liquid metal systems for which
the wetting transitions exists but since the metastable l/l/v line
it falls below the liquidus line it is hidden.

The largest thickness of the gravitationally limited Bi rich
wetting layers that formed at the free surface at
(l/l/v)-coexistence were typically of the order of $\approx
50\AA$. As a result of the fortuitous balance between the surface
potential that favors the Bi-rich Phase and the gravitational
potential which favors the lighter Ga-rich phase the interface
dividing the two coexisting phases is sufficiently close to the
free surface that x-ray reflectivity was sensitive to its
microscopic structure.  We interpreted these measurements in terms
a Landau type of square gradient phenomenological theory from
which it was possible to extract the sole free parameter of that
model, i.e. the influence parameter $\kappa$. To the best of our
knowledge this is the first time that such a parameter has been
directly determined from measurement.  Furthermore, it was
possible to make a  distinction between the intrinsic width of the
interface and the additional contributions to the width from
thermal capillary wave features fluctuations. On making using of
available bulk thermodynamic data, along with the extracted
influence parameter, we were able to calculate the experimentally
difficult to get (l/l) interfacial tension for a wide temperature
range. As a test of our methods we performed the same calculation
of the (l/l) interfacial tension for Ga-Pb with its analogous
miscibility. We assumed that the influence parameter, $\kappa$, as
determined from our measurements on the (l/l) interfacial
structure in Ga-Bi, was also applicable to Ga-Pb and, using the
thermochemical data sets available for Ga-Pb, we obtained good
agreement between our calculated values and macroscopic
measurements of the (l/l) interfacial tension. This suggests that
the value of the influence parameter extracted in this study might
provide a reasonable value for the larger class binary metallic
alloys. If this proves to be true, then the value determined here
can be used, along with the surface potential, to predict wetting
properties of this larger class of systems.

\acknowledgements We thank Prof. S. Dietrich and Prof. B. I.
Halperin for helpful discussions. This work is supported by U.S.
DOE Grant No. DE-FG02-88-ER45379, National Science Foundation
Grant DMR-0124936, and the U.S.-Israel Binational Science
Foundation, Jerusalem. BNL is supported by U.S. DOE Contract No.
DE-AC02-98CH10886. Patrick Huber acknowledges support from the
Deutsche Forschungsgemeinschaft.

\appendix*
\section{bulk thermodynamics} Here\footnote{Note unless it is is explicitly indicated otherwise, in this chapter
''$T$'' denotes the absolute temperature in Kelvin units.}, we
focus on the part of the phase diagram dominated by the
miscibility gap and the monotectic point M. In a liquid-liquid
(l/l) equilibrium of binary systems, consisting of two components
(Ga and Bi resp.), at atmospheric pressure p, temperature T, the
compositions $c_{\rm I}$, $c_{\rm II}$ of two coexisting bulk
phases (I and II) are given by the thermodynamic conditions:
\begin{eqnarray}
\mu_{\rm Ga}^{\rm I}(T)=\mu_{\rm Ga}^{\rm II}(T)~~~~ \mu_{Bi}^{\rm
I}(T)=\mu_{\rm Bi}^{\rm II}(T) \label{eq:muequilib}
\end{eqnarray}
where $\mu$ is the chemical potential. The experimental conditions
of the experiments that we are going to present are constant
temperature, pressure and a fixed number of moles of each
component, therefore we have to resort to the Gibbs free energy
$G$ in order to describe the thermodynamic equilibrium:
\begin{eqnarray}
G&=&U-TS+PV \\
G&=&n\;g=n_{\rm Ga}\;\mu_{\rm Ga}\,+\,n_{\rm Bi}\;\mu_{\rm Bi}
\end{eqnarray}
Since the total number of particles in mol $n=n_{\rm Ga}+n_{\rm
Bi}$ is constant and only two components have to be considered,
the system can be described with a molar Gibbs Free Energy g(c,T)
that depends on only on the molar fraction of one component:
\begin{eqnarray}
G\,&=&\, n\;(c\, \mu_{\rm Ga}\,+\,(1-c)\,\mu_{\rm Bi}\,)\\
\Rightarrow g(c,T)\,&=&\,c\,\mu_{\rm Ga}+(1-c)\,\mu_{\rm Bi}
\end{eqnarray}
With this notation the phase equilibrium conditions (equation
(\ref{eq:muequilib}) transform into the following two equations:
\begin{eqnarray}
\label{commontangent}
 \frac{\partial g}{\partial c} |c^{\rm I}
&=&\frac{\partial g}{\partial c} |c^{\rm II}\\
 \frac{\partial g}{\partial c}|c^{\rm I}&=&\frac
 {g(c^{I})-g(c^{II})}{c^{I}-c^{II}}
\label{eq:commontangent}
\end{eqnarray}
Having a model for the free energy of a binary sytem, it is
possible, then, to calculate $c_{\rm I}$ and $c_{\rm II}$ of the
coexisting phases by solving of equation (\ref{eq:commontangent}).
The standard approach to describe a miscibility gap in a binary
demixing system is the regular solution model for the Gibbs free
energy \cite{Atkins2001}. The resulting miscibility gap has a
symmetric shape and is centered around the consolute point $c_{\rm
crit} = 0.5$ in the ($c$,$T$)-plane, quite in contrast to the
miscibility gap of Ga-Bi with its asymmetric shape, centered
around $c_{\rm crit}$=0.7 in the ($c$,$T$)-plane. Therefore, it is
necessary to resort to an \textit{extended} regular solution
model. Relying on data sets for Ga-Bi from the Calphad
initiative\cite{Kaufman2001}, we use such a model based on
Redlich-Kister polynoms $L_{\rm \nu}(T)$\cite{Kattner1997,
Lukas1982} and express $g(c,T)$ in the following way:
\begin{center}
\begin{eqnarray}
g(c,T)&=&c\cdot g_{\rm 0}(Ga)(c,T)+(1-c)\cdot g_{\rm 0}(Bi)\nonumber\\
      & &+R\cdot T \left [\,c\;ln(c)\;+\;(1-c)\,ln(1-c)\right ] \label{eq:gERSM}\\
      & &+\;\Delta g_{\rm mix}(c,T)\nonumber\\
 \Delta g_{\rm
mix}(c,T)~&=&~c(c-1)~\sum_{\rm \nu=1}^{5}L_{\rm
\nu}(T)(1-2c)^{\nu}\nonumber
\end{eqnarray}
\end{center}
The first two terms correspond to the Gibbs energy of a mechanical
mixture of the constituents; the second term corresponds to the
entropy of mixing for an ideal solution, and the third term,
$\Delta g_{\rm xs}$ is the so-called excess term. Here, it is
represented by an extension of the usual regular solution
expression, i.e. a sum of Redlich-Kister polynomials, $L_{\rm
\nu}(T)$, as listed in Table \ref{tab:RedKisterPolys}.

\begin{table}
\squeezetable
\begin{ruledtabular}
\begin{tabular}{c|c}
  $\nu$ & Redlich-Kister polynomial L$_{\rm \nu}(T)$\\
  \hline
  0 & 80000-3389+T \\
  \hline
  1 & -4868-2.4342$\cdot T$ \\
  \hline
  2 & -10375-14.127$\cdot T$ \\
  \hline
  3 & -4339.3 \\
  \hline
  4 & 2653-9.41$\cdot$T\\
  \hline
  5 & -2364\\
\end{tabular}
\end{ruledtabular}
\caption{\label{tab:RedKisterPolys}Redlich-Kister polynomials used
to model the (l/l) miscibility gap of Ga-Bi\cite{Kaufman2001,
Kattner1997}.}
\end{table}

By solving the nonlinear equations (\ref{eq:commontangent})
applied on $g(c,T)$ for the temperature range $T_{\rm M} < T <
262^{\circ}C$ we could calculate the binodal coexistence line as
plotted in Fig. \ref{fig:GaBiPDcT}. The agreement of the
calculated phase boundaries with Predel's measured phase
boundaries is excellent. Particularly, the measured consolute
point C ($T_C=262^{\circ}C$, $c_{\rm crit}=0.7$ is nicely
reproduced by the calculated critical values $T_{\rm C}=262.8°C$
and $c=0.701$. Furthermore, the knowledge of $g(c,T)$ allows us to
extrapolate the binodal lines below $T_{\rm M}$ into the region of
metastable (l/l) coexistence, and, hence, to get information on
the energetics of the metastable Ga-rich, Bi-rich phase resp.
Similar to the procedure presented above, we also calculated the
phase boundaries below $T_{\rm M}$. Here, we used data sets for
the Gibbs free energy of pure solid Bi, which coexists for
$T<T_{\rm M}$ with a Ga-rich liquid.



\begin{references}
\bibitem{Cahn1977}  J. W. Cahn, J. Chem. Phys. \textbf{66}, 3667
(1977).
\bibitem{Ebner1977}  C. Ebner and W. F. Saam, Phys. Rev.
Lett. \textbf{38}, 1486 (1977).
\bibitem{DeGennes1985}  P. G. de Gennes, Rev. Mod. Phys. \textbf{57} ,827(1985).
\bibitem{Sullivan1986} D. E. Sullivan and M. M. T. da Gama, \textit{Fluid Interfacial Phenomena}, edited by C. A. Croxton (Wiley, New York), (1986).
\bibitem{Dietrich1988} S. Dietrich, in C. Domb and J. L. Lebowitz (Eds.), Phase
Trans. and Crit. Phen., Vol. 12, Acad. Press, NY, 1988.
\bibitem{Law2001}  B. M. Law, Progress in Surface Science \textbf{66}, 159
(2001).
\bibitem{Bonn2001} D. Bonn, D. Ross, Rep. Prog. Phys. \textbf{64}, 1085 (2001).
\bibitem{Plech2000}  A. Plech, U. Klemradt, M. Huber, J. Peisl, Europhys. Lett. \textbf{49},
583(2000).
\bibitem{Tostmann2000}  H. Tostmann, E. DiMasi, O. G. Shpyrko, P. S. Pershan, B. M. Ocko, M. Deutsch, Phys. Rev. Lett. \textbf{84}, 4385
(2000).
\bibitem{Schlossman2002} M. L. Schlossman, Curr. Opinion Coll.\&Interf. Sc. \textbf{7}, 235
(2002).
\bibitem{Chatain1996}  D. Chatain and P. Wynblatt, Surface Science \textbf{345}, 85
(1996).
\bibitem{Shim2001} H. Shim, P. Wynblatt, D. Chatain, Surface Science \textbf{476}, L273 (2001).
\bibitem{Nattland1995}  D. Nattland, P. D. Poh, S. C. M\"{u}ller, W. Freyland, J. Phys. C \textbf{7}, L457
(1995).
\bibitem{Predel1960}  P. Predel, Z. f. Phys. Chemie Neue Folge \textbf{24}, 206
(1960).
\bibitem{Kaufman2001}  The calculations of the phase boundaries rely on data
sets for Ga-Bi from the CalPhaD initiative (Larry Kaufman (MIT)),
which have been refined in order to more accurately reproduce
Predel's phase diagram \cite{Predel1960}.
\bibitem{Perepezko1982} J. H. Perepezko, C. Galang, K. P. Cooper in G. E. Rindone (ed.), Materials Processing in the Reduced Gravity
Environment of Space, Elsevier (Amsterdam) p 491 (1982).
\bibitem{Lei1996}  N. Lei, Z. Q. Huang, and S. A. Rice, J. Chem. Phys. \textbf{104}, 4802 (1996).
\bibitem{Dietrich1997}  S. Dietrich and M. Schick, Surface Science \textbf{382}, 178
(1997).
\bibitem{Huber2002} P. Huber, O. G. Shpyrko, P. S. Pershan, B. M. Ocko, E. DiMasi, M. Deutsch,
 Phys. Rev. Lett. \textbf{89}, 035502 (2002).
\bibitem{Regan1997} M. J. Regan, P. S. Pershan, O. M. Magnussen, B. M. Ocko, M. Deutsch, L. E. Berman,
Phys. Rev. B \textbf{55}, 15874 (1997).
\bibitem{Pershan1984}  P. S. Pershan and J. Als-Nielsen, Phys. Rev. Lett. \textbf{52}, 759
(1984).
\bibitem{Pershan1994}  P. S. Pershan, Phys. Rev. E \textbf{50}, 2369
(1994).
\bibitem{Tolan1998} M. Tolan, \textit{X-Ray Scattering from Soft-Matter Thin Films},
Springer Tracts in Modern Physics \textbf{148} (1998).
\bibitem{Parratt1954} L. G. Parratt, Phys. Rev. \textbf{95} 358 (1954).
\bibitem{Lipowsky1986} R. Lipowsky, D. A. Huse, Phys. Rev. Lett. \textbf{57}, 353 (1986).
\bibitem{Wu1986} X. L. Wu, M. Schlossman, C. Franck, Phys. Rev. B \textbf{33}, 402 (1986).
\bibitem{Fenistein2002} D. Fenistein, D. Bonn, S. Rafai, G. H. Wegdam, J. Meunier, A.O. Parry, M.M.T. da Gama,
 Phys. Rev. Lett. \textbf{89}, 096101 (2002).
\bibitem{Davis1996} H. T. Davis, \textit{Statistical Mechanics of Phases, Interfaces, and Thin Films}, Wiley-VCH, NY (1996).
\bibitem{Schmid2001} F. Schmid, N. B. Wilding, Phys. Rev. E \textbf{63}, 1201 (2001).
\bibitem{Plech2001}  A. Plech, U. Klemradt, J. Peisl, J.Phys. C \textbf{13}, 5563 (2001).
\bibitem{Bonn2002} D. Fenistein, D. Bonn, S. Rafai, G. H. Wegdam, J. Meunier, A. O. Parry, M. M. T.
 da Gama, Phys. Rev. Lett. \textbf{89}, 096101 (2002).
\bibitem{Ashcroft1991} N. W. Ashcroft, Phil. Trans. R. Soc.
Lond. A \textbf{334}, 407 (1991).
\bibitem{Pluis1989} B. Pluis, T. N. Taylor, D. Frenkel, J.F. van
der Veen, Phys. Rev. B \textbf{40} 1353 (1989).
\bibitem{VanderWaals1894} J. D. van der Waals, Z. Phys. Chem.
\textbf{13} (1894).
\bibitem{Cahn1958} J. W. Cahn, J. E. Hilliard, J. Chem. Phys. \textbf{28}, 258 (1958).
\bibitem{Safran1994} S. A. Safran, \textit{Statistical Thermodynamics of Surfaces,
Interfaces, and Membranes}, Frontiers in Physics, ed. David Pines,
Westview Press (1994).
\bibitem{Enders1998} S. Enders, K. Quitzsch, Langmuir \textbf{14}, 4606 (1998).
\bibitem{Buff1965} F. P. Buff, R. A. Lovett , F. H. Stillinger, Phys. Rev. Lett. \textbf{15} 621
(1965).
\bibitem{Rowlinson1982} J. S. Rowlinson, B. Widom, \textit{Molecular Theory
of Capillary}, Clarendon Press, Oxford (1982).
\bibitem{Ocko1994}  B. M. Ocko, X. Z. Wu, E. B. Sirota, S. K. Sinha, M. Deutsch,
Phys. Rev. Lett. \textbf{72}, 242 (1994).
\bibitem{Heilmann2001} R. K. Heilmann, M. Fukuto, P. S. Pershan, Phys. Rev. B \textbf{63}, 5405 (2001).
\bibitem{Braslau1988} A. Braslau, P. S. Pershan, G. Swislow, B. M. Ocko, J. Als-Nielsen,
 Phys. Rev. A \textbf{38}, 2457 (1988).
\bibitem{Wadewitz1999} T. Wadewitz, J. Winkelmann, Phys. Chem. Chem. Phys. \textbf{1}, 3335 (1999).
\bibitem{Evans1981} R. Evans, Mol. Physs \textbf{42}, 1169 (1981).
\bibitem{Daillant1999} J. Daillant, A. Gibaud, \textit{X-Ray and
Neutron Reflectivity: Principles and Applications}, Lecture Notes
in Physics, Springer (1999).
\bibitem{Stauffer1972}  D. Stauffer, M. Wortis, M. Ferer, Phys. Rev. Lett. \textbf{29}, 345
(1972).
\bibitem{Kreuser1993}  H. Kreuser and D. Woermann, J. Chem. Phys. \textbf{98}, 7655
(1993).
\bibitem{Merkwitz1998} M. Merkwitz, J. Weise, K. Thriemer, W. Hoyer, Zeitschrift f\"ur Metallkunde \textbf{89}, 247 (1998).
\bibitem{Wynblatt1998} P. Wynblatt, A. Saul, D. Chatain, Acta
Metallurgica \textbf{46}, 2337 (1998).
\bibitem{Ansara1991} I. Ansara, F. Ajersch J. Phase Equilibria
\textbf{12} 73 (1991).
\bibitem{Cornelisse1997} P. M. W. Cornelisse, C. J. Peters, J. de Arons, J. Chem. Phys. \textbf{106}, 9820 (1997).
\bibitem{Fisk1969} S. Fisk, B. Widom, J. Chem. Phys. \textbf{50}, 3219 (1969).
\bibitem{Sengers1989} J. V. Sengers, J. M. J. VanLeeuwen, Phys. Rev. A \textbf{39}, 6346 (1989).
\bibitem{Schick1990}  M. Schick, in J. Charvolin, J. F. Joanny, J. Zinn-Justin (Eds.), Liquids at Interfaces, Vol. XLVIII, p.
419.
\bibitem{Brezin1983} E. Brezin, B. I. Halperin, S. Leibler, Phys. Rev. Lett. \textbf{50}, 1387 (1983).
\bibitem{Kroll1985} D. M. Kroll, R. Lipowsky, R. K. P. Zia, Phys. Rev. B \textbf{32}, 1862 (1985.)
\bibitem{Khokonov1974} K. B. Khokonov, S. N. Zadumkin, Sov. Electrochem. \textbf{10},
865 (1974).
\bibitem {Ayyad2002a} A. H. Ayyad, W. Freyland, Surf. Science \textbf{506}, 1
(2002).
\bibitem {Ayyad2002b} A. H. Ayyad, I. Mechdiev, W. Freyland, Chem. Phys.
Lett. \textbf{359}, 326 (2002).
\bibitem{Pandit1983} R. Pandit, M.E. Fisher, Phys. Rev. Lett.
\textbf{51} 1772 (1983).
\bibitem{Zimmerli1992} G. Zimmerli, M. H. W. Chan, Phys. Rev. B \textbf{45}, 9347 (1992).
\bibitem{Turchanin2001} A. Turchanin, D. Nattland, W. Freyland, Chem. Phys. Lett. \textbf{337}, 5 (2001).
\bibitem{Turchanin2002} A. Turchanin, W. Freyland, D. Nattland, Phys. Chem. Chem. Phys. \textbf{4}, 647 (2002).
\bibitem{Earnshaw1992} J. C. Earnshaw, C. J. Hughes, Phys. Rev. A \textbf{46}, R4494
(1992).
\bibitem{Wu1993} X. Z. Wu, R. Sikorski, S. K. Sinha, B. M. Ocko, M. Deutsch,
Phys. Rev. Lett. \textbf{70}, 958 (1993).
\bibitem{Atkins2001} P. W. Atkins, \textit{Physical Chemistry}, W.H.
Freeman (2001).
\bibitem{Kattner1997} U. R. Kattner, JOM-J. Minerals Metals\& Mat. Soc. \textbf{49}, 14 (1997).
\bibitem{Lukas1982} H. L. Lukas, J. Weiss, and E.-Th. Henig, CALPHAD, \textbf{6} 229
(1982).
\bibitem{Maede1996} D. Maede, B. S. Haran, R. E. White, MapleTech \textbf{3}, 85 (1996).

\end{references}
\end{document}